\documentstyle{article}
 \newcommand {\nc}{\newcommand}
 \nc{\eq}{\begin{equation}}
 \nc{\en}{\end{equation}}
 \def\complex{{\cal C}}
 \def\Alg{{\cal A}}
 \def\module{{\cal M}}

 \def\tld{\tilde}
 \def\pr{\prime}
 
 \def\dlt{\delta}

 \newtheorem{definition}{Definition}
 
 \newtheorem{theorem}{Theorem}
 
 \newtheorem{corollary}{Corollary}

 \def\prf{{\bf{Proof:}}\\}
 \def\endprf{${\bf{\Box}}$\\}
 \title{A No-Go Theorem for the Compatibility between Involutions of the First Order Differentials on a Lattice
 and the Continuum Limit}
 \author{Jian Dai, Xing-Chang Song\\
 Institute of Theoretical Physics, School of Physics, Peking University\\
 Beijing, P. R. China, 100871\\
 jdai@mail.phy.pku.edu.cn,
 songxc@ibm320h.phy.pku.edu.cn}
 \date{May 6th, 2001; Revision: Oct 15th, 2001}
 
\begin{document}
  \begin{titlepage}
  \maketitle
  \begin{abstract}
   We prove that the following three properties can not match each other on a lattice,
   that differentials of coordinate functions
   are algebraically dependent to their involutive conjugates,
   that the involution on a lattice is an antihomomorphism and
   that differential calculus has a natural continuum limit.\\


  \end{abstract}
  \end{titlepage}
  \section{Introductions}
  {\it Why lattices as noncommutative spaces?} To the majority
  of the high-energy physics community, it is no more than a {\it technique} of a universal regulator for the
  nonperturbative definition of QCD to discretize the space-time to be a
  lattice \cite{Creutz}; however, it is shown to be fruitful at a more fundamental level in
  \cite{DMS1} where Wilson action for lattice gauge field
  is recovered by the virtue of noncommutative calculus over commutative algebras and in \cite{L}\cite{FSW}
  where axial anomaly in $U(1)$ lattice gauge theory is analyzed adopting noncommutative differential
  calculus, that a lattice is treated as an independent geometric
  object whose geometry is characterized by a {\it noncommutative
  relation}
  \eq\label{eq2}
   [x_\mu, dx_\nu]=-\dlt_{\mu\nu}a_\mu dx_\nu,
  \en
  which is
  a deformation of the ordinary commutation relation between coordinates functions
  $x_\mu$ and their differentials $dx_\mu$,
  subjected to a set of lattice constants $a_\mu$.
  Note that on one hand Eq.~(\ref{eq2}) is a special case in the category of differential
  calculi over commutative associative algebras \cite{BDM1} and that on the other hand this equation
  illustrates intuitively the {\it bi-local} nature of differentials on lattices.
  This philosophy that a lattice is a simple model of {\it noncommutative geometry} will be
  bore in this work. \\

  {\it Why involutions on lattices?} An {\it involution} $\ast$ is an {\it antihomomorphism}
  of a complex algebra $\Alg$,
  fulfilling the requirements that
  \eq\label{Inv}
   a^{\ast\ast}=a, (a+b)^\ast=a^\ast +b^\ast, (\lambda a)^\ast=\bar{\lambda}a^\ast, (ab)^\ast=b^\ast a^\ast
  \en
  where $a, b\in\Alg$, $\lambda\in\complex$ and $\bar{\lambda}$ denotes the complex
  conjugation; it is a generalization of complex conjugation and
  the hermitian conjugation of complex matrices. In any complex
  regime, it is utilized to define {\it real objects}, for example
  {\it compact real forms} in complex Lie algebras. Moreover physically it
  is necessary from the silent feature that in a generic gauge theory a gauge potential is
  expressed as a {\it real} 1-form and that the dynamics of gauge fields is controlled by
  Lagrangian $\langle F^\ast|F\rangle$ where $F$ is the strength 2-form and $\langle|\rangle$ is the
  contraction of differential forms,
  to extend an involution over an algebra into the space of differential forms according to the same
  conditions in Eq.~(\ref{Inv}).
  There is a {\it canonical involution} on complex functions on a lattice, defined by
  pointwise complex conjugation; the problem to extend it into the
  differential algebra on this lattice will be addressed below.\\

  {\it Consistency between differentials and involutions.}
  However, a na\"{\i}ve extension of the canonical involution on a lattice
  will encounter inconsistency immediately; in fact, act any such {\it would-be} involution
  to both sides of
  Eq.~(\ref{eq2}) in terms of Eq.~(\ref{Inv}),
  \[
   dx_\nu^\ast x_\mu^\ast-x_\mu^\ast dx_\nu^\ast=-\delta_{\mu\nu}a_\mu^\ast dx_\nu^\ast,
  \]
  under the {\it natural} assumptions that $x_\mu$ and $a_\mu$ are real, and
  that $dx_\mu^\ast$ is linear dependent on $dx_\mu$ one has
  \[
   [x_\mu, dx_\nu]=\dlt_{\mu\nu}a_\mu dx_\nu,
  \]
  which is {\it contradictive} to Eq.~(\ref{eq2})! This paradox implies some {\it natural}
  assumptions for geometric and algebraic structures over lattices are incompatible,
  like antihomomorphic rule and ordinary continuum limit for for involution, the algebraic
  dependency of differential and its image under involution; we will formulate this observation
  into more rigid scrutinies and prove a
  {\it no-go theorem} for an involution on lattice.
  In our understanding though the proof is simple, the conclusion
  is highly nontrivial and very inspiring to understand the problem of {\it continuum limit} in
  lattice field theory. In section~\ref{Sect:2} noncommutative
  geometry of lattices as well as other mathematical facts are prepared; the {\it no-go} theorem is
  stated and proved in section~\ref{Sect:3}; some discussions are
  put in section~\ref{Sect:4}.
  \section{Noncommutative geometry of lattices}\label{Sect:2}
  First an algebraic concept has to be introduced for further application.
  \begin{definition}
   Let $\Alg$ be an algebra over a generic field and $\module$ is
   a bimodule over $\Alg$. $\module$ is left(right) finitely-represented if
   $\module$ is able to be expressed as a finitely-generated left(right) $\Alg$-module.
  \end{definition}
  Only D-dimensional hyper-cubic lattices will be considered
  below, with lattice constant along $\mu$ direction being written
  as $a_\mu$. $\Alg$ is specified to be the algebra of functions on this lattice over complex numbers;
  coordinate functions $x_\mu$ $\mu=1...D$ are
  valued in integers. The above-mentioned inconsistency is
  exposed at the level of first order differential, thus only the space of
  {\it 1-forms} will be considered.
  \begin{definition} A First order differential calculus over
  $\Alg$ is a pair $(\module,d)$ in which $\module$ is an $\Alg$-bimodule and
  $d\in Hom_\complex(\Alg, \module)$ satisfying Leibnitz rule
  \eq\label{Leibnitz}
   d(ff^\pr)=d(f)f^\pr +fd(f^\pr), \forall f,f^\pr\in\Alg.
  \en
  \end{definition}
  {\it Continuum limit}, short as {\it C.L.}, is referred to $max(\{a_\mu\})\rightarrow
  0$. To have a correct {\it C.L.}, $\module$ is required to be
  generated algebraically by images $d(x_\mu)$ and to be {\it left finitely-represented} accordingly to
  Eq.~(\ref{eq2}) being rewritten as
  \eq\label{eq2:2}
   [x_\mu, d(x_\nu)]=-\dlt_{\mu\nu}a_\mu d(x_\nu);
  \en
  hence {\it C.L.} is also a {\it commutative limit};
  Eq.~(\ref{eq2:2}) is also referred as {\it structure equation}
  of $\module$.
  For the mathematical
  rigidity, suppose that there is no two-sided ideal in $\Alg$ except $\{0\}$ annihilating $\module$.
  Under above assumptions, the {\it first order differential} $d$
  can be parameterized by
  \[
   d(f)=\nabla_\mu(f)d(x_\mu)
  \]
  for any $f\in\Alg$ with coefficient functions $\nabla_\mu(f)$.
  \begin{corollary}(Deformed Leibnitz rule on $\Alg$)
   \eq\label{Leibnitz:f}
    \nabla_\mu(ff^\pr)=\nabla_\mu(f)\tld{T}_\mu(f^\pr)+f\nabla_\mu(f^\pr),\forall
    f,f^\pr\in\Alg
   \en
   where $\tld{T}_\mu$ are translations acting on $\Alg$ by
   $\tld{T}_\mu(f)(p)=f(T_\mu(p))$, $x_\nu(T_\mu(p))=x_\nu(p)+\dlt_{\mu\nu}a_\mu$ for any $f$ in
   $\Alg$ and $p$ in the lattice.
  \end{corollary}
  \prf
    It can be checked by using
    Eqs.~(\ref{Leibnitz})(\ref{eq2:2}).\\
  \endprf
  An {\it involution on $\Alg$} is a specification of Eq.~(\ref{Inv})
  \[
   f^{\ast\ast}=f,
   (f+f^\pr)^\ast=f^\ast +f^{\pr\ast},
   (ff^\pr)^\ast=f^\ast f^{\pr\ast},
   (\lambda f)^\ast=\bar{\lambda}f^\ast
  \]
  for all $\lambda\in\complex$, $f, f^\pr\in\Alg$; also
  for a correct {\it C.L.}, involution on $\Alg$ is required
  to be pointwisely convergent to canonical involution
  \eq\label{CL:1}
   |f^\ast(p)-\overline{f(p)}|\stackrel{C.L.}{\rightarrow} o(1)
  \en
  for all $p$. An {\it involution on $\module$} is an antihomomorphism of
  $\module$ satisfying
  \eq\label{Inv:M}
   v^{\ast\ast}=v,
   (\lambda v)^\ast=\bar{\lambda}v^\ast,
   (v+v^\pr)^\ast=v^\ast +v^{\pr\ast},
   (fv)^\ast =v^\ast f^\ast, (vf)^\ast=f^\ast v^\ast
  \en
  for all $\lambda\in\complex$, $f\in\Alg$, $v, v^\pr\in\module$.
  Because of $\module$ being left finitely-represented and the
  structure equation Eq.~(\ref{eq2:2}), any involution on $\module$ can be characterized by a collection
  of coefficient functions
  \[
   d(x_\mu)^\ast=\iota_{\mu\nu}d(x_\nu)
  \]
  Note that the commutation diagram relation between $d\circ \ast$
  and $\ast\circ d$ is irrelevant to our purpose and that
  $d(x_\mu)^\ast$ will be understood henceforth as
  $[d(x_\mu)]^\ast$.
  \section{No-go theorem}\label{Sect:3}
  \begin{theorem}
   There does not exist an involution on $\module$, which has a
   natural continuum limit
   \eq\label{CL:2}
    |\iota_{\mu\nu}(p)-\dlt_{\mu\nu}|\stackrel{C.L.}{\longrightarrow}
    o(1)
   \en
  \end{theorem}
  \prf
   Suppose that there exists such an involution $\ast$ that satisfies Eq.~(\ref{CL:2}).
   The structure equation Eq.~(\ref{eq2:2}) can be
   rewritten as
   \eq\label{SE}
    d(x_\nu)x_\mu=(x_\mu+\dlt_{\mu\nu}a_\mu)d(x_\nu)
   \en
   Apply $\ast$ on the both sides of Eq.~(\ref{SE}), and note the definition of involution Eq.~(\ref{Inv:M}),
   \eq\label{eq}
    x_\mu^\ast d(x_\nu)^\ast=d(x_\nu)^\ast(x_\mu^\ast +\dlt_{\mu\nu}a_\mu)
   \en
   The behaviors of involution under {\it C.L.} i.e.
   Eqs.~(\ref{CL:1})(\ref{CL:2}) can be described as
   \eq\label{eq4}
    x_\mu^\ast=x_\mu+\alpha_{\mu\sigma}(x)a_\sigma,
    d(x_\mu)^\ast=d(x_\mu)+\beta_{\mu\rho\sigma}(x)a_\rho d(x_\sigma)
   \en
   in which $\alpha_{\mu\nu}(x)$ are restrictions of $C^1$ functions on the lattice
   and $\beta_{\lambda\mu\nu}(x)$ are restrictions of bounded
   functions on the lattice. Substitute Eq.(\ref{eq4}) into
   Eq.(\ref{eq}), use the structure equation Eq.~(\ref{eq2:2})
   again and remember that {\it no two-sided ideal of $\Alg$ except $\{0\}$ annihilates
   $\module$},
   \eq\label{eq0}
    (\dlt_{\nu\lambda}+\beta_{\nu\rho\lambda}a_\rho)(\dlt_{\mu\nu}a_\mu +\dlt_{\mu\lambda}a_\mu +\tld{\nabla}_\lambda
    (\alpha_{\mu\sigma})a_\sigma)=0
   \en
   in which $\tld{\nabla}_\mu:=\tld{T}_\mu-Id$. Note that this
   definition of {\it finite differences} fulfills the deformed Leibnitz
   rule in Eq.~(\ref{Leibnitz:f}). Now sum Eq.~(\ref{eq0}) over
   $\mu$;
   \eq\label{Judg}
    (\dlt_{\nu\lambda}+\beta_{\nu\rho\lambda}a_\rho)(a_\nu +a_\lambda +\tld{\nabla}_\lambda
    (\hat{\alpha}_\sigma)a_\sigma)=0
   \en
   in which $\hat{\alpha}_\sigma:=\sum_{\mu=1}^D
   \alpha_{\mu\sigma}$; so $\hat{\alpha}_\sigma$ are still
   restrictions of $C^1$ functions. Consider these special cases in
   Eq.~(\ref{Judg}) that $\nu=\lambda$;
   \eq\label{Judg_1}
    (1+\beta_{\nu\rho\nu}a_\rho)(2a_\nu +\tld{\nabla}_\nu
    (\hat{\alpha}_\sigma)a_\sigma)=0.
   \en
   And pick out a special limit procedure that $a_\mu=a\rightarrow
   0$ for all $\mu$; then for any $\nu$ Eq.~(\ref{Judg_1}) implies
   that
   \eq\label{Last}
    (1+a\hat{\beta_\nu})(2 +\tld{\nabla}_\nu
    (\check{\alpha}))=0
   \en
   where $\hat{\beta_\nu}=\sum_{\rho=1}^D\beta_{\nu\rho\nu}$ and
   $\check{\alpha}=\sum_{\sigma=1}^D\hat{\alpha}_\sigma$.
   Therefore $\hat{\beta_\nu}$ continues to be restrictions of bounded
   functions and $\check{\alpha}$ is a restriction of a $C^1$
   function; thus the left-hand side of Eq.~(\ref{Last}) will be
   greater than $1$ when $a$ is small enough, which makes
   Eq.(\ref{Last}) fails to be an identity! \\

   The {\it no-go} theorem follows this contradiction.
  \endprf
  \section{Discussions}\label{Sect:4}
  In \cite{DMS1},
  the above paradox is avoided by defining $\ast$ to be a
  homomorphism instead; this solution can not be generalized to
  the case where $\Alg$ becomes noncommutative however.
  In \cite{F1}, consistent involution on abelian discrete groups, with lattice being
  taken as a class of special cases, is defined to be
  $f^\ast (g)=\overline{f(-g)}$; it violates the requirement of
  correct continuum limit. In \cite{DS1}, $dx_\mu$ and $dx_\mu^\ast$ are algebraically independent
  generators of first order differential forms; this so-called {\it nearest symmetric reduction} has also been mentioned
  as an example in \cite{D0}. \\

  Due to the above-proved no-go theorem,
  coordinate functions have to be suppose to be algebraic independent
  to their involutive images, if antihomomorphic rule
  and continuum limit of involution are regarded as being more natural and more necessary, which in physics
  implies that a connection
  1-form, thus a gauge field, has two components along one
  direction! If a lattice formalism of field theory is taken to be
  a {\it microscopic description} of our continuum world, any
  inferences of this doubling of degrees of freedom in gauge
  theory are very interesting.
  \section*{Acknowledgements}
    This work was supported by Climb-Up (Pan Deng) Project of
    Department of Science and Technology in China, Chinese
    National Science Foundation and Doctoral Programme Foundation
    of Institution of Higher Education in China.
  

\begin{thebibliography}{120}
   \bibitem{Creutz} M. Creutz, {\it Quarks, gluons, and lattices},
   Cambridge University Press (New York) 1983.
   \bibitem{DMS1}
   A. Dimakis, F. M\"{u}ller-Hoissen, T. Striker,
   ``Non-commutative differential calculus and lattice gauge
   theory'',
   J. Phys. A: Math. Gen. {\bf 26} (1993) 1927-1949.
   \bibitem{L}
   M. L\"{u}scher,
   ``Topology and the axial anomaly in abelian lattice gauge
   theories'',
   Nucl. Phys. B{\bf 538} (1999) 515-529,
   hep-lat/9808021.
   \bibitem{FSW}
   T. Fujiwara, H. Suzuki, K. Wu,
   ``Non-commutative Differential Calculus and the Axial Anomaly in Abelian Lattice Gauge
   Theories'',
   Nucl. Phys. B{\bf 569} (2000) 643-660, hep-lat/9906015;
   {\it ibid},
   ``Axial Anomaly in Lattice Abelian Gauge Theory in Arbitrary
   Dimensions'',
   Phys. Lett. B{\bf 463} (1999) 63-68, hep-lat/9906016;
   {\it ibid},
   ``Application of Noncommutative Differential Geometry on Lattice to
   Anomaly'',
   IU-MSTP/37, hep-lat/9910030.
   \bibitem{F1}
   B. Feng,
   ``Differential Calculus on Discrete Groups and its Application
   in Physics'',
   MS Disertation (1997, Peking University).
   \bibitem{DS1}
   J. Dai, X-C. Song,
   ``Noncommutative Geometry of Lattice and Staggered Fermions'',
   Phys. Lett. B{\bf 508}(2001)385-391,
   hep-th/0101130.
   \bibitem{D0}
   A. Dimakis, F. M\"{u}ller-Hoissen,
   ``Differential Calculus and Discrete Structures'',
   GOET-TP 98/93,
   in the proceedings of the International Symposium ``Generalized Symmetries in Physics'' (ASI Clausthal, 1993),
   hep-th/9401150.
  \end{thebibliography}
 \end{document}